\documentclass[twocolumn,showpacs,preprintnumbers,amsmath,amssymb]{revtex4}

\usepackage{graphicx}
\usepackage{dcolumn}
\usepackage{bm}

\begin{document}

\preprint{APS/123-QED}

\title{Experiments with low energy ion beam transport \\  into toroidal magnetic fields}

\author{N. Joshi}
\email{joshi@iap.uni-frankfurt.de}
\author{M. Droba}
\author{O. Meusel}%
\author{U. Ratzinger}
\affiliation{%
Institute for Applied Physics (IAP), Goethe University \\
600438 Frankfurt, Germany 
}%


\date{January 20, 2010}

\begin{abstract}
The stellarator-type storage ring for accumulation of multi- Ampere proton and ion beams with energies in the range of $100~AkeV$ to $1~AMeV$ is designed at Frankfurt university.
The main idea for beam confinement with high transversal momentum acceptance was presented in EPAC2006.
This ring is typically suited for experiments in plasma physics and nuclear astrophysics.
The accumulator ring with a closed longitudinal magnetic field is foreseen with a strength up to $6-8~T$.
The experiments with two room temperature 30 degree toroids are needed.
The beam transport experiments in toroidal magnetic fields were first described in EPAC2008 within the framework of a proposed low energy ion storage ring.
The test setup aims on developing a ring injection system with two beam lines representing the main beam line and the injection line.
The primary beam line for the experiments was installed and successfully commissioned in 2009.
A special diagnostics probe for \textit{"in situ"} ion beam detection was installed.
This modular technique allows online diagnostics of the ion beam along the beam path.
In this paper, we present new results on beam transport experiments and discuss transport and transverse beam injection properties of that system.

\end{abstract}

\maketitle

\section{\label{sec:level1}Introduction\protect\\
 }

At Frankfurt University a storage ring for low energy high current density ion beams is proposed \cite{Droba_1}.
A longitudinal magnetic field component will provide a quite homogenous transverse beam focusing conditions along the whole structure.
The main advantage of a stellarator type ring against the conventional one is the higher transverse momentum acceptance which opens a possibility to accumulate high beam currents.
The planned ring comprises of curved sectors with longitudinal magnetic field to form the Figure-8 geometry. 
The continual and coupled longitudinal magnetic field should provide focusing and guiding forces.
In curved magnetic fields, charged particles tend to drift leaving the desired trajectories. 
To compensate the drift arising from curved magnetic fields a twisted geometry is preferred which looks like a classical stellarator with a Figure-8 shape.
High magnetic field of about $6-8~T$ produced by superconducting coils is desired to minimize drift forces and to achieve high current densities.

\begin{figure}
\includegraphics[width=80mm]{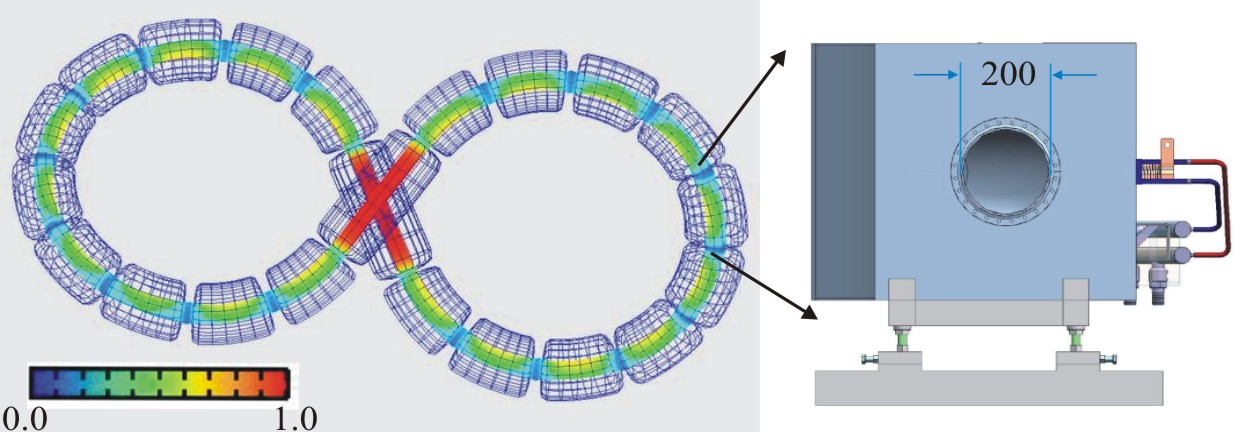}
\caption{\label{fig:01}
An example of ring formed with toroidal segments depicting single magnetic surface. 
Magnetic field strength (normalized) is colour coded. Room temperature prototype segment is shown on the right.
}
\end{figure}

This ring can be built using multiple toroidal segments. 
Due to the 3-dimensional geometry the magnetic field lines do not close in a single turn around, but tend to form a surface, known as a magnetic flux surface.
The torsion of the field lines, called rotational transform, is assumed to increase the stability for confined charged particles. 
One example of  such a magnetic surface is shown in Fig.~\ref{fig:01}.

The room temperature experiments scaled down to $0.6~ T$ were planned to investigate the drift dynamics, space charge effects, beam diagnostics and multiturn beam injection system \cite{Droba_2}.
The properties of prototype magnets are stated in Table~\ref{tab:table1}.
The measurements from these experiments are compared with the numerical model which will be used to design the injection system.

\section{\label{sec:level2}Drift dynamics in curved magnetic fields\protect\\
 }

The dynamics of charged particle beams in magnetic fields is characterized by gyro motion.
The beam size undergoes through periodic maxima and minima in longitudinal magnetic field when injected from field free region.
In curved magnetic fields the ion beam is guided on a circular path, additionally dominated by the drift motion, namely $\mathbf{R} \times \mathbf{B}$ drift due to curved magnetic field lines, $\nabla\mathbf{B} $ drift due to the inhomogeneous field and $\mathbf{E} \times \mathbf{B}$ drift due to the space charge \cite{Chen}.

The curvature drift is given by,

\begin{eqnarray}
\mathbf{v_{R \times B}}=\frac{mv_{\|}^2}{qB^2}~\frac{\mathbf{R} \times \mathbf{B} } {R^2}
\label{eq:02}.
\end{eqnarray}

This equation implies that a particle when injected into the magnetic field will experience a drift force along the normal formed by the radius vector $\mathbf{R}$ and magnetic field $\mathbf{B}$.
For a proton beam with an energy of $10~keV$,  injected in the magnetic field of $0.6~T$ and $R=1.3~m$, it is $\mathbf{v_{R \times B}}=25.70\times10^3~m/s$.
Along with the time of flight one can calculate this drift to be $12.6~mm$ along the beam path over an arc distance of $680~mm$.

The $\nabla\mathbf{B}$ drift arises due to the difference of coil density on the inner side and outer side of the toroid.
The drift velocity due to this inhomogeneous field is given by,

\begin{eqnarray}
\mathbf {v_{\nabla \mathbf {B}}}= \pm \frac{1}{2} v_\bot r_L \frac{\mathbf B \times \nabla \mathbf{B}}{B^2}
\label{eq:03},
\end{eqnarray}
where $r_L$ is Larmor radius.
In the case of used toroidal segments the magnetic field gradient is $\nabla \mathbf{B}=0.4~T/m$.
For a proton beam with a transversal velocity spread up to $1.66\times10^5~m/s$ (corresponding to $v_{\perp}/v_\parallel=120~mrad$ measured from experiments), the drift velocity can be maximum $\mathbf {v_{\nabla \mathbf {B}}}=158~m/s$.

An additional drift arises due to the self electric field of the beam. The crossed terms in the Lorentz force equation implies the particles will experience additional drift velocity given by,

\begin{eqnarray}
\mathbf{v_{E \times B}}=\frac{\mathbf{E} \times \mathbf{B}}{B^2} 
\label{eq:04}.
\end{eqnarray}

The homogeneous ion beam has a maximum coulomb force on the boundary.
Thus a cross product of radial component of electric field ($E_r$) and longitudinal magnetic field ($B_\zeta$) gives rise to rotation of the beam around its axis.
For a proton beam ($r=15~mm$) with current $I=3.0~mA$ at $10~keV$ energy this drift velocity is $\mathbf{v_{E \times B}}=4.31\times10^3~m/s$.

Thus we conclude in our case the curvature drift is the most dominating over the others.

\section{\label{sec:level3}Simulation tool - Toroidal Beam Transport (TBT) \protect\\
 }

A computer code (TBT) was written to simulate the dynamics of particles in given fields.
TBT can simulate a realistic external fields e.g. magnetic field from coil or electric field from electrodes.
The Biot-savart law is used to calculate the magnetic field at toroidal grid points from current carrying pancake coils as a source.
The standard toroidal coordinate system ($r,\theta, \zeta$) was used  \cite{Balescu}.

A space-charge subroutine uses a Particle in Cell (PIC) method in 3-dimensional toroidal geometry \cite{Hockney,Birdsall}.
A first order charge distribution scheme uses relative toroidal volume as a weighing factor.
The potential at grid points is solved using the Poisson equation.
An iterative method Bi-Conjugate Gradient Stabilized (BiCGS) is used to solve resulting two dimensional matrix with $N_1\times N_2 \times N_3$ linear equations, where $N$'s represent the number of grid points in each direction \cite{wolfram}.

\begin{figure}
\vspace{5mm}
\includegraphics[width=80mm]{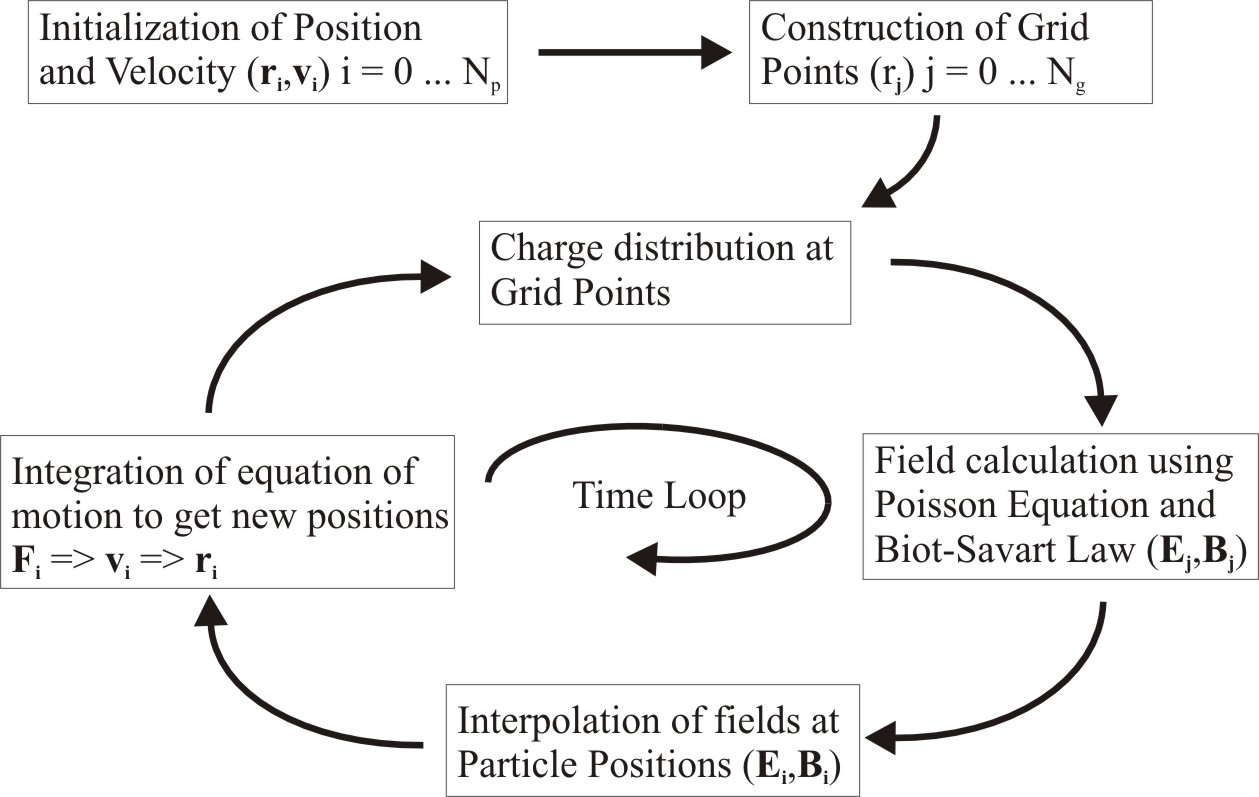}
\caption{\label{fig:02}Flow chart of the TBT algorithm.}
\end{figure}

A middle step Finite-Difference Time-Domain scheme calculates position advance in the time domain. 
The flowchart in Fig.~\ref{fig:02} shows the complete algorithm.

\begin{figure}[!h]
\includegraphics[width=80mm]{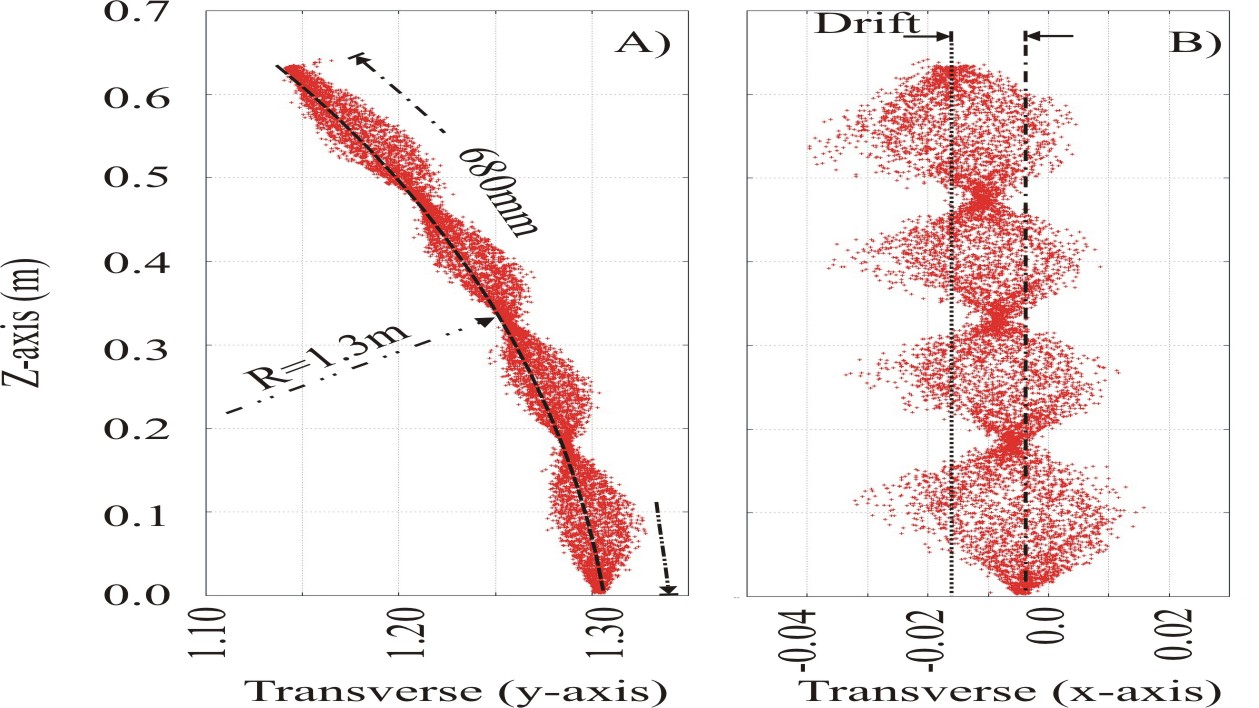}
\caption{\label{fig:03}Projections of proton beam envelope with energy $10~keV$ transported into the magnetic field of $0.6~T$. A) Projection on the $y-z$ plane showing curved beam, B) Projection on the $x-z$ plane showing the beam drift.}
\end{figure}

For example, consider a parallel beam of proton at an energy of $10~keV$ and a nominal beam current of $2~mA$ injected into the single toroidal segments.
Fig.~\ref{fig:03} shows the beam envelope is projected on the $y-z$ and the $x-z$ planes.
For a theoretical analysis a homogeneous input distribution of particle with a beam radius of $8~mm$ and angle spread of $80~mrad$ was defined.

The graph on the left shows beam curved with $30$ arc degree angle.
And the graph on the right shows beam drifted vertically in the negative x-direction from central axis about $13~mm$.
Multiple beam waist can be located along the beam path.
Along with the time of flight the number of beam waist during the transport can be calculated. 

\section{\label{sec:level4}Experiments \protect\\
 }
 
The experimental setup was designed by taking into account three main aspects, ion beam production, matching into toroidal field and diagnostics.
A volume type ion source was constructed to produce light ion species.
The high voltage terminal supports up to $20~kV$.
Up to $20~keV$ beams can be extracted using triode extractor.
Pure $He^+$ beam can be extracted from $He-$plasma, whereas $H^+$, $H_{2}^+$ and $H_{3}^+$ - beams can be formed by a proper choice of $H_2-$plasma parameters \cite{Joshi_2}.
The ion source was mounted on the vacuum chamber with a Faraday cup.
The solenoid was installed downstream for beam matching.
It provides a maximum field of $0.72~T$ on the axis.

\begin{table}[!h]
\caption{\label{tab:table1}Experimental parameters}
\begin{ruledtabular}
\begin{tabular}{lc}
Aspect&Quantity\\
\hline
Ion source & Hot filament volume type \\
Extraction & Triode extraction\\
Ion species & $He^+, H^+, H_{2}^+,H_{3}^+$ \\
Beam energy & $20~keV~max$ \\
$He^+$ & $2.0~mA~@~10~keV$\\
$H_{3}^+$ & $ \sim 95\%\sim 3.0~mA~@~10~keV$\\
$H_{2}^+$ & $ \sim 91\%\sim 2.8~mA~@~10~keV$\\
$H^+$ & $ \sim 45\% \sim 2.8~mA~@~10~keV$\\
\hline
Solenoid &  \\
No. of windings & $280$ \\
Maximum on axis magnetic field & $0.72~T$ \\
Maximum voltage and current & $32.5~V, ~360~A$ \\
Length & $250~mm$ \\
Diameter of aperture& $106~mm$ \\
Magnetic Shielding & present \\ 
\hline
Toroid &  \\
No. of windings & $33 \times 24$\\
Maximum on axis magnetic field & $0.6~T$ \\
Maximum Voltage and Current & $140~V,480~A$ \\
Major Radius $R_0$ & $1300~mm$ \\
Arc angle & $30\,^\circ$ \\
Arc length & $680~mm$ \\
Diameter of aperture & $200~mm$ \\
Magnetic Shielding & absent \\ 
Cooling water & $70~l/min$ \\
Weight & $1050~kg$ \\
\end{tabular}
\end{ruledtabular}
\end{table}

An optical assembly used for the beam diagnostics, consists of a phosphor screen and a digital camera. 
It was installed in the flange downstream of toroidal segment with a ring electrode in front of it.
The potential up to $\pm1.2~kV$  can be put to this electrode to attract or repel any secondary electrons.
Phosphor screen was composed of $(Zn,Cd)S:Ag$, known as $P20$, emits a light within a range of $470~nm-670~nm$ and with a peak emission at $550~nm$,  a yellow-green colour. 
The screen had a diameter of about $128~mm$.
A digital camera produces $8-bit$ image in the $jpeg$ format.
A separate image manipulation routine was written to analyse the image produced.

\subsection{\label{sec:level5}Beam Transport in a Single Toroidal Segment \protect\\
 }
 
In the first stage the dynamics in single toroidal segment was studied. 
The optical assembly was fixed downstream of the magnet with a proper magnetic shielding for the camera.

\begin{figure}[!h]
\includegraphics[width=80mm]{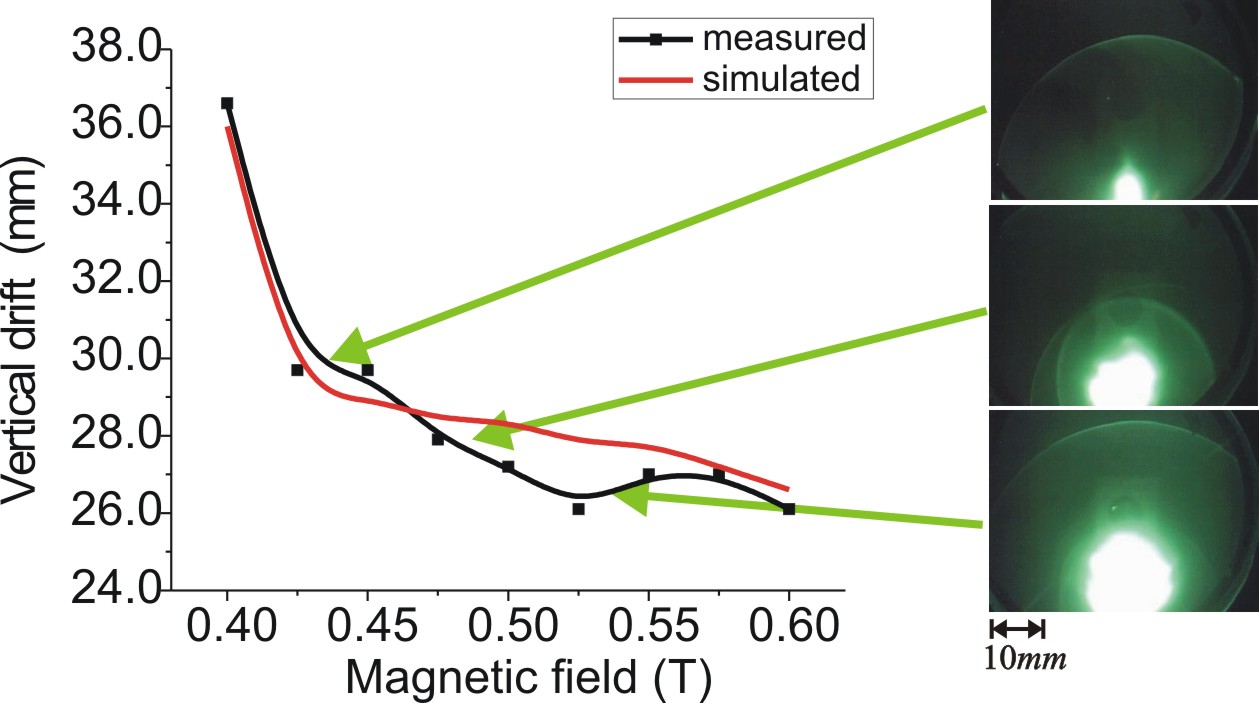}
\caption{\label{fig:04}A proton beam at the energy of $12~keV$ detected downstream of the toroid for three different values of magnetic field. The measured vertical drifted position as a function of B-field is plotted and compared with the simulation.}
\end{figure}

The beam drifts were measured with this method as a function of the magnetic field. 
In the case of proton beam, the beam is composed of three different ion species (see Table~\ref{tab:table1}), the drifted position of the beam do not follow simple $1/B$ dependence.
Fig.~\ref{fig:04} shows that results are in the good agreement with simulations.
In this case, a proton beam at the energy of $12~keV$ was injected in the toroidal magnetic field.
The drifted position was calculated with respect to geometrical axis calibrated before measurements.

\begin{figure}[!h]
\includegraphics[width=80mm]{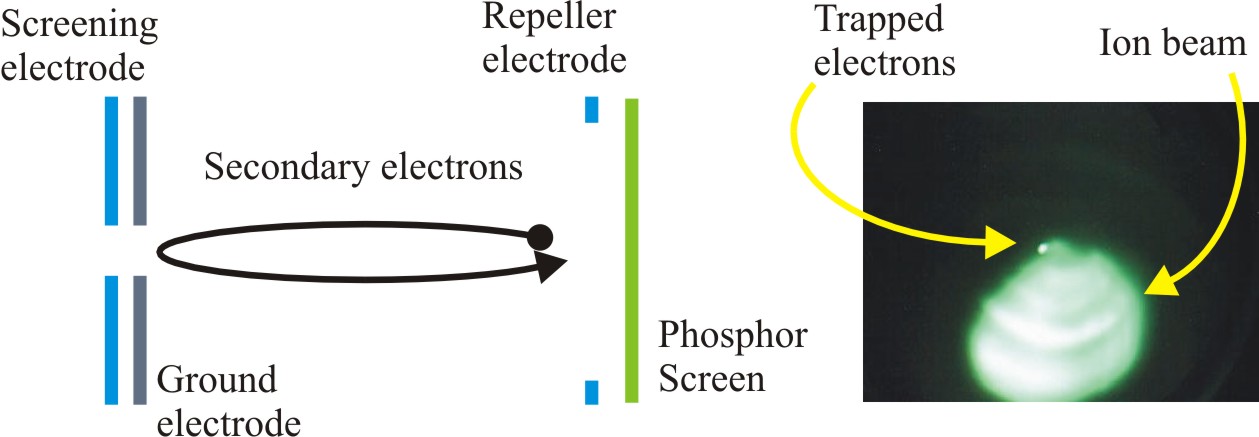}
\caption{\label{fig:05}Perturbed negative potential from screening electrode of the ion source and the potential due to the repeller electrode biased negative form a longitudinal trap for electrons. Secondary electrons produced at the center of the phosphor screen giving a magnetic center. On the right central electrons can be seen along with a drifted ion beam.}
\end{figure}

The electrons are produced along the beam line due to beam loss (secondary electrons) or ion collision with rest gas atoms.
It is assumed that some of the electrons are longitudinally confined in the transport channel.
Along the beam line, perturbed negative potential from screening electrode of the ion source and the potential due to the repeller electrode when biased negative form a longitudinal trap for electrons.
These electrons can be observed to produce a small spot at the center of the screen (see Fig.~\ref{fig:05}).

These electrons may have energy in the range of $5-10~eV$.
Since electrons have $1/1846$ times the mass of proton, experience very low curvature drift less than $0.1~mm$.
This is practically negligible, giving a magnetic center of the system.

\subsection{\label{sec:level6}Diagnostics: Pigging Technique\protect\\
}

\begin{figure}[!h]
\includegraphics[width=80mm]{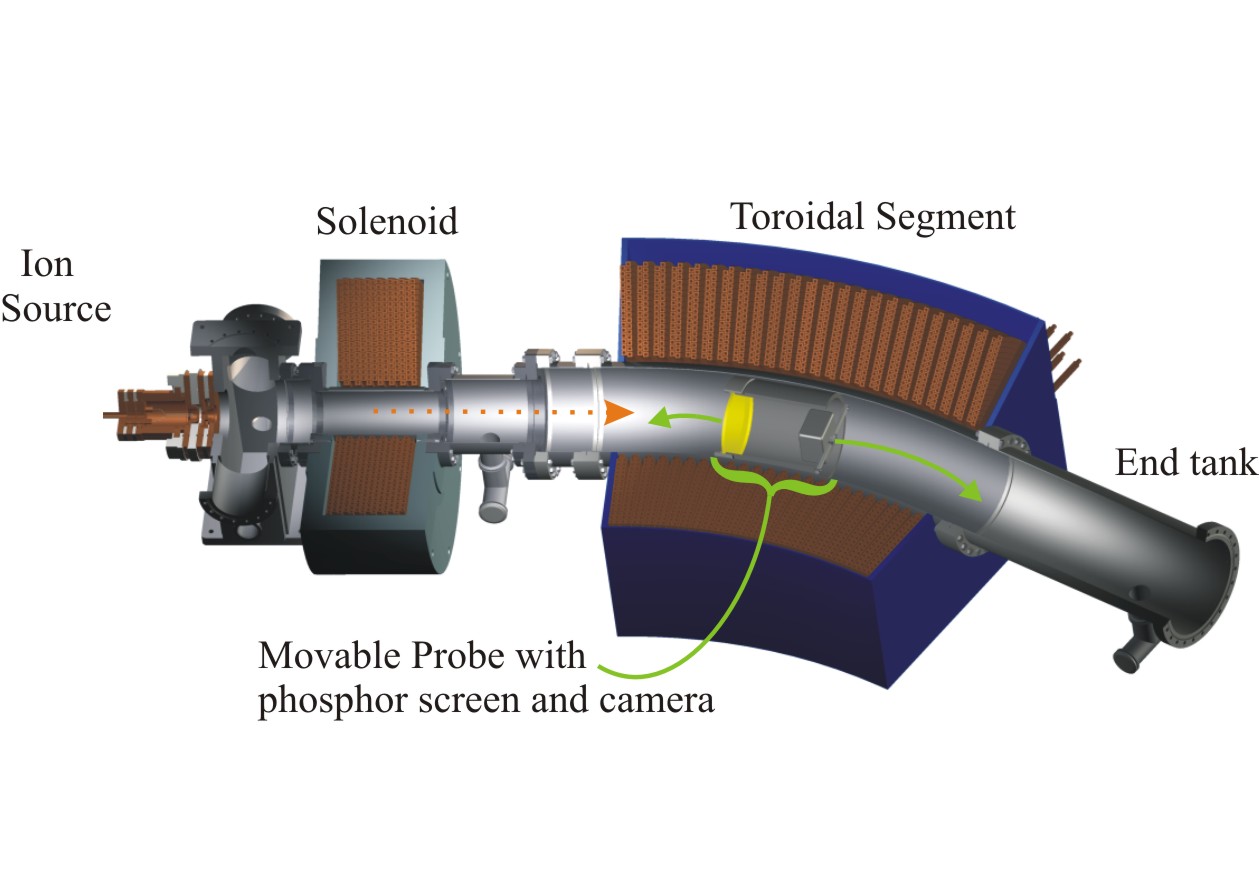}
\caption{\label{fig:06}The experimental setup showing ion source, solenoid, toroidal sector magnet and the movable probe.}
\end{figure}

An innovative "Pigging technique" was developed for the beam diagnostics.
The optical assembly was upgraded to manoeuvre the optical probe along the beam path.
The phosphor screen and the camera were installed on a movable cylinder (see Fig.~\ref{fig:06}).
This assembly is able to sustain a magnetic field up to $0.6~T$ and high vacuum conditions.
A repeller ring ($\pm 1.2~ kV$) in front of the screen was fixed.
This detection probe gives an opportunity to detect a transversal beam profile along the longitudinal axis.

\begin{figure}[!h]
\includegraphics[width=80mm]{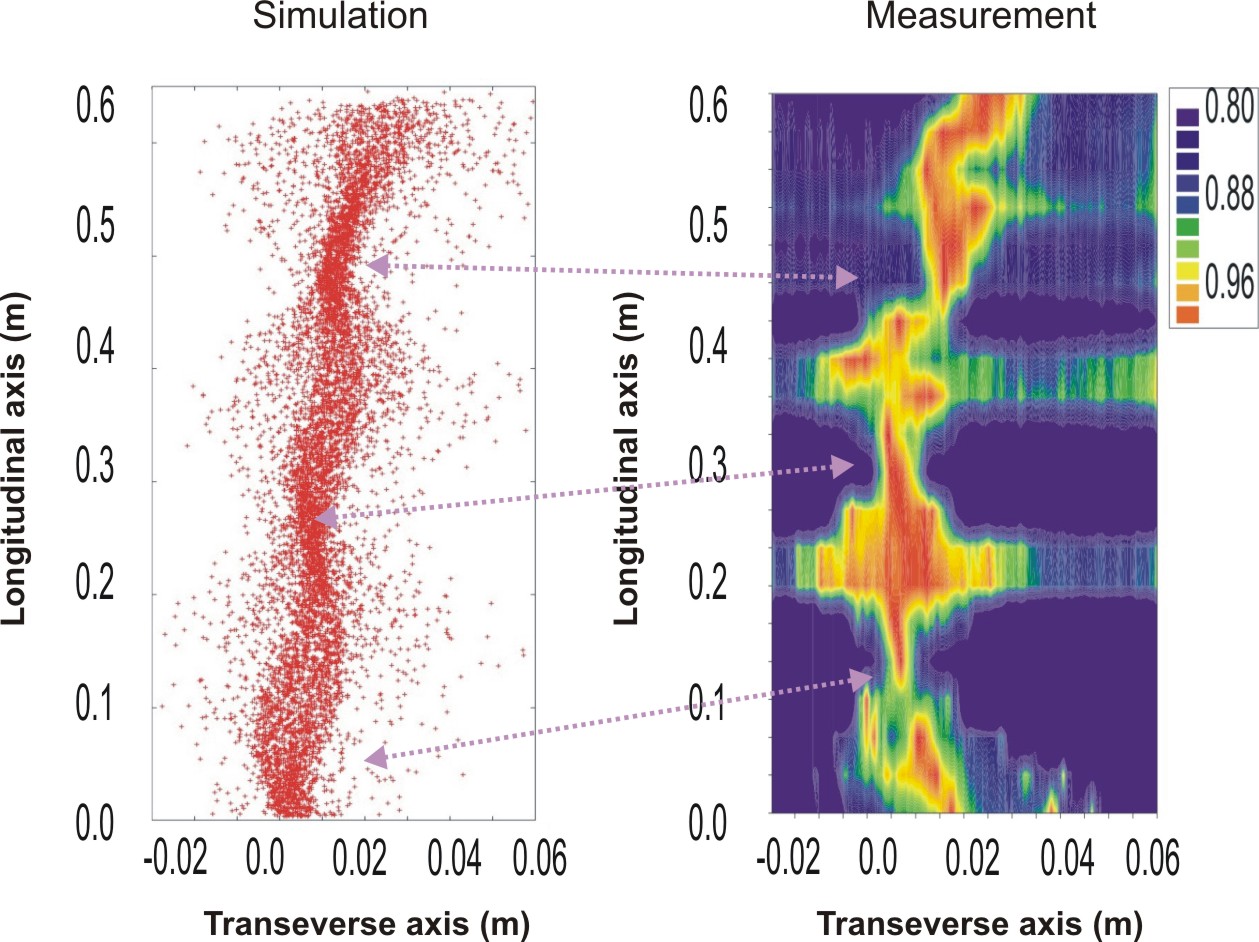}
\caption{\label{fig:07}Transversal beam profile of the ion beam ($He^+$-beam at $5~ keV$) along the beam path, simulations (left) and measurements (right). The locations of the beam waist are indicated with arrows.}
\end{figure}

Fig.~\ref{fig:07} compares the simulation results with a transversal profile measurement along the longitudinal axis.
$He^+$-beam with energy of $5~keV$ was injected into the toroid with a magnetic field of $0.6~T$.
As the beam is transported into the segment it shows periodic beam waists.
The number of beam waists along the path can be calculated from beam energy and field strengths.
Simulations show a good agreement with measurements with respect to drifted position and number of beam waists.
The errors are caused mostly due to inaccuracy and low resolution in the probe positioning along the longitudinal axis.
When the beam is diverging the spot size is larger than the probe diameter.
The secondary electron production is also high at these positions.
This leads to band like structure in the measurements.
The deviation in the location of the beam waist from the simulation results can also be noticed.
This is presumed due to electron entrapment between solenoid and toroid.
Early experiments using Langmuir probe have shown the presence of electron cloud with a density of about $10^{11}/m^{3}$ in this region.
The longitudinal focus shift due to a plasma column is known \cite{Meusel}.

\subsection{\label{sec:level7}Beam Transport in Coupled Segments \protect\\
 }
  \begin{figure*}
\includegraphics[width=160mm]{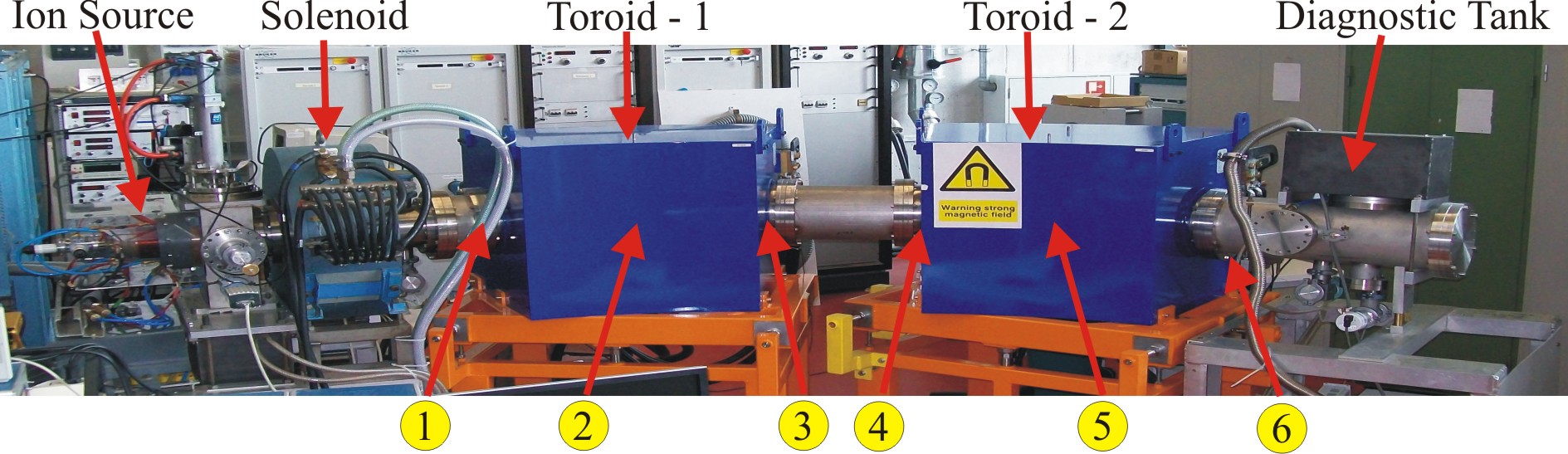}
\caption{\label{fig:08}The setup for the beam transport experiments along two coupled toroidal segments.}
\end{figure*}

Two segments were coupled in the next stage (see Fig.~\ref{fig:08}).
The diagnostic probe can be moved along the beam path over the total distance of $1760~mm$ from position-1 to position-6 as marked in the photograph. 
This kind of setup provides a possibility for various experiments not only in terms of parameters but also with respect to the geometrical arrangement of magnet components.
In the first step a "Straight Section", an intermediate separation with a length of $400~mm$ was chosen.
The whole arrangement forms a $60\,^\circ$ part of the ring. 
Along the longitudinal axis the magnetic field forms a ripple like structure.

\begin{figure}[!h]
\includegraphics[width=80mm,height=40mm]{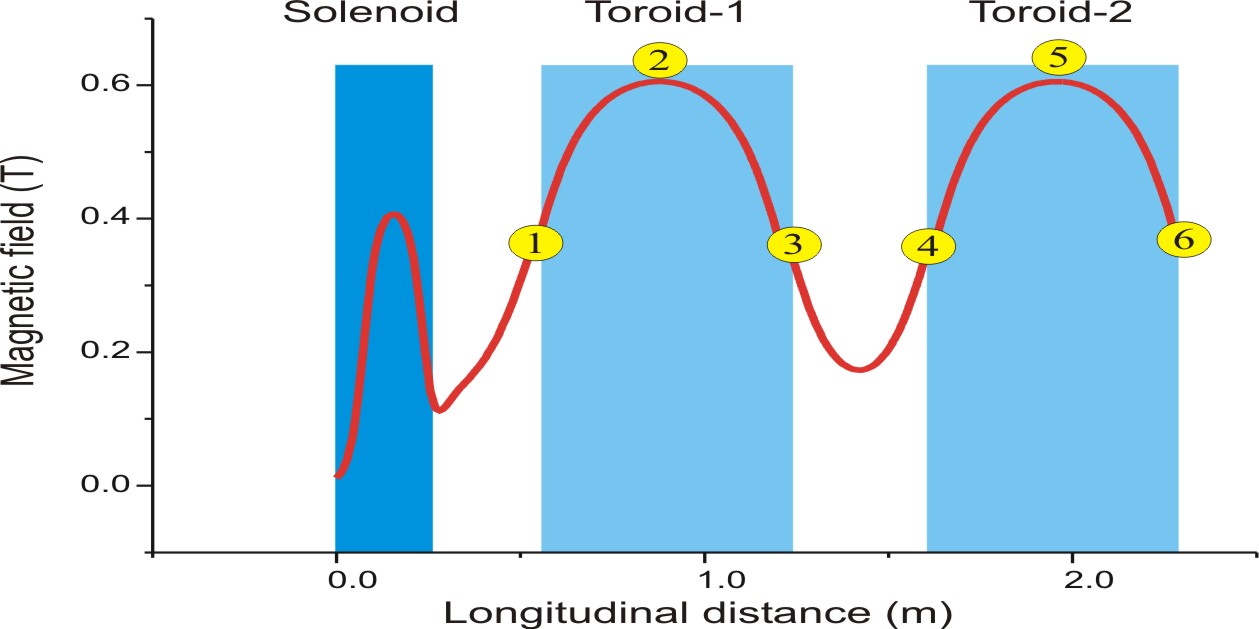}
\caption{\label{fig:09}Magnetic field along the longitudinal axis.}
\end{figure}

The magnetic field drops down to $0.2~T$ constituting about $30\%$ from maximum field strength of $0.6~T$ (see Fig.~\ref{fig:09}).
There have been indications of electron cloud entrapment between positions 3-4 due to mirror configuration of the magnetic field, according to the observations from the beam transport experiments.

\begin{figure}[!h]
\vspace{0.5cm}
\includegraphics[width=80mm]{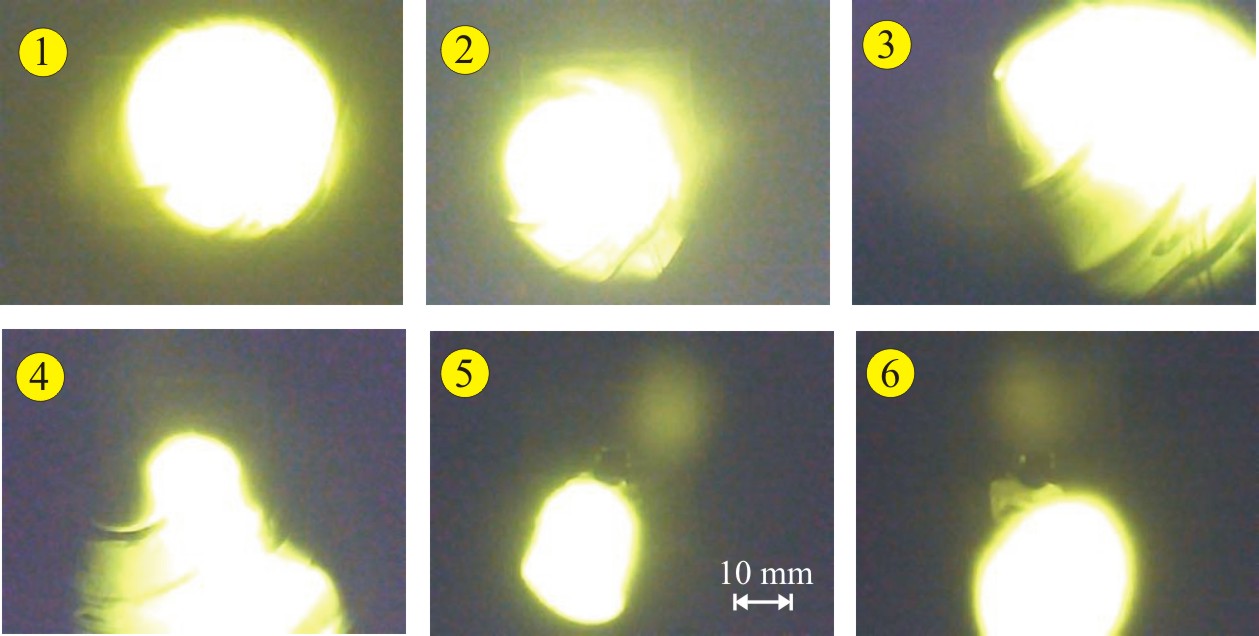}
\caption{\label{fig:10}Measurement of proton beam with an energy of $9.70~keV$ at different positions along the beam path when transported through two coupled segments. }
\end{figure}

Fig.~\ref{fig:10} shows a low energy proton beam ($9.70~keV$, $ 2.2~ mA$) detected at various positions along the beam path.
Position-1 shows clearly defined circular beam spot at the entrance of the first toroid.
It can be seen as quite homogenous beam with a well defined boundary.
Position 2 shows a beam that is drifted vertically downward due to the curvature drift.
From position 3 to position 4 the magnetic field drops down drastically and the field lines expand in the space.
The ions following the field lines are forced toward the vessel wall. 
This leads to beam losses in the straight section.
Thus the beam at these positions is not well defined rather influenced by the secondary electrons.
At position 5 the beam is recaptured into the acceptance of the second toroid.
While the position 6 shows, the beam detected just at the exit of the second toroid.

Overall, vertically drifted position along the beam path (1-6) can be seen due to the curvature drift.
The space charge effects due to electrons on the ion beam will be investigated in detail especially in the ripple region.

\begin{figure}[!h]
\vspace{0.5cm}
\includegraphics[width=80mm,height=80mm]{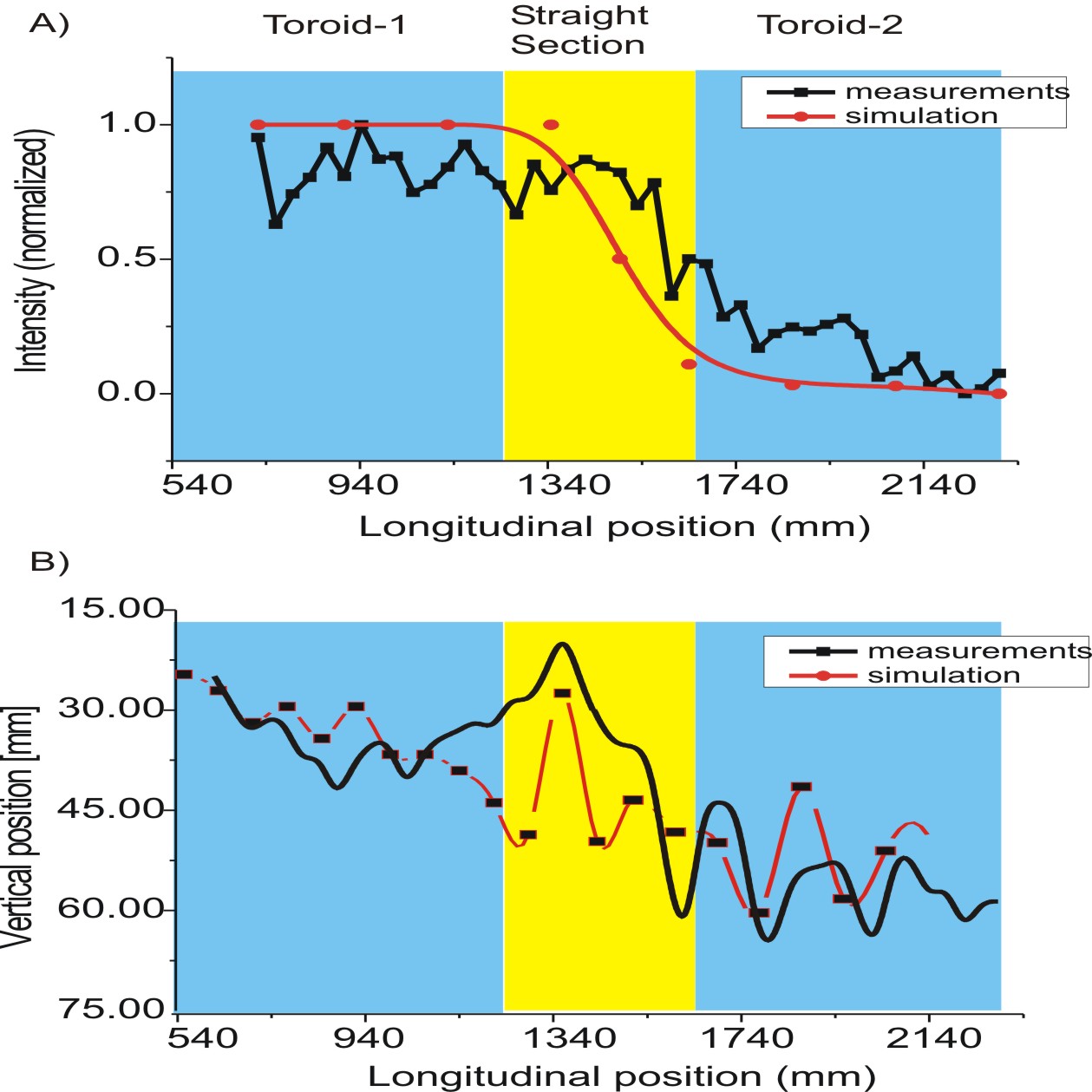}
\caption{\label{fig:11}A) Transmission along the longitudinal axis simulated and measured; B) Vertical position of beam plotted along longitudinal path.}
\end{figure}

Fig.~\ref{fig:11} A) shows the transmission (detected by optical screen intensity) as a function of longitudinal position of the probe. Maximum losses are seen in the straight section due to weak coupling.
Fig.~\ref{fig:11} B) shows the vertically drifted position of an ion beam. The measurement is in reasonable agreement with the analytical value resulting in $32~mm$. The simulation shows discrepancy near the exit of first toroid. But the presence of high peak in the straight section and large oscillations in the second toroid can be readily spotted.
The differences between the simulation and measurement are blamed on rotation of the probe, lack of correct positioning system and camera sensitivity.

In further experiments the effect of separation distance between two segments will be investigated.
Theoretically zero distance is the best case for beam transport.
But some space is required for injection experiment to install second beam line.
The best suitable case of $300~mm$ as predicted by simulation will be chosen for the final setup.
The beam loss is expected to reduce by about factor $3$ comparative to the earlier case.
This improvement is sufficient to show the effects of separation.
The drift compensation experiments are planned by installation of second segment in the opposite direction forming an S - shape.

\section{\label{sec:level8}Injection experiments \protect\\
 }

\begin{figure}[!h]
\includegraphics[width=80mm]{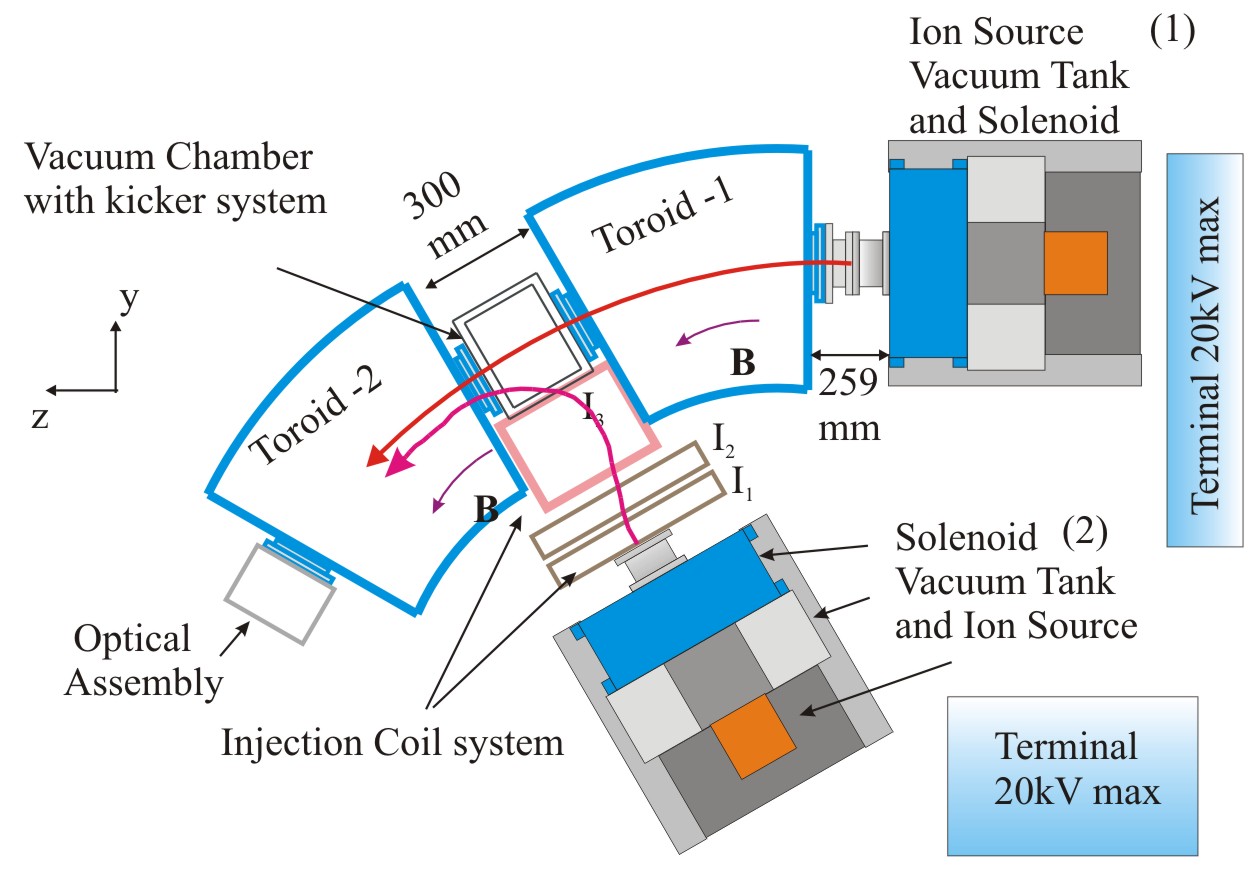}
\caption{\label{fig:12}Setup for the beam injection experiments.}
\end{figure}

The challenges in injecting charged particles from field free region into the confinement region has been addressed and investigated especially using electron beams \cite{Clark,Berkery}.
The experimental setup has been designed to study injection experiments for such a type of accumulator ring.
Two segments will be arranged on circular arc to form a $60\,^{\circ}$ part of storage ring (see Fig.~\ref{fig:12} ).
The injection line will be installed between two of these sector magnets.
Identical ion source and solenoid will be installed to upgrade the existing beam line.

The beam optics for injection is calculated with the simulation tool TBT.
A special magnetic field combination is designed to inflect the proton beam into the acceptance of Toroid-2.
The 3d-phase space portrait is used to describe beam quality.
To extract the information of guidance of beam along magnetic field lines a parameter \textit{velocity ratio} ($v_{\lambda}$) was defined as,

\begin{eqnarray} 
velocity~ ratio=v_{\lambda}=\left( \frac {v_{\perp}}{v_{\parallel}} \right)_{\mathbf{B}}
\label{eq:06},
\end{eqnarray}

where $ v_{\parallel} $ is defined as a velocity component parallel to the magnetic field at a particular position and magnetic field strength distribution.
This ratio is also referred as \textit{pitch} in some of the publications.
Three dimensional map then can be produced by plotting $v_{\lambda}$ as a function of relevant space coordinate (e.g. $x-z$ at the input of the injection system and $x-z$ at the output, the exit of Toroid-2).

A kicker system which make use of $\mathbf{E} \times \mathbf{B}$  drift force would move the injected beam onto the main (ring) field lines.
The whole geometry is optimized for a proton beam with the energy of $10~keV$.

\begin{figure}[!h]
\vspace{5mm}
\includegraphics[width=80mm]{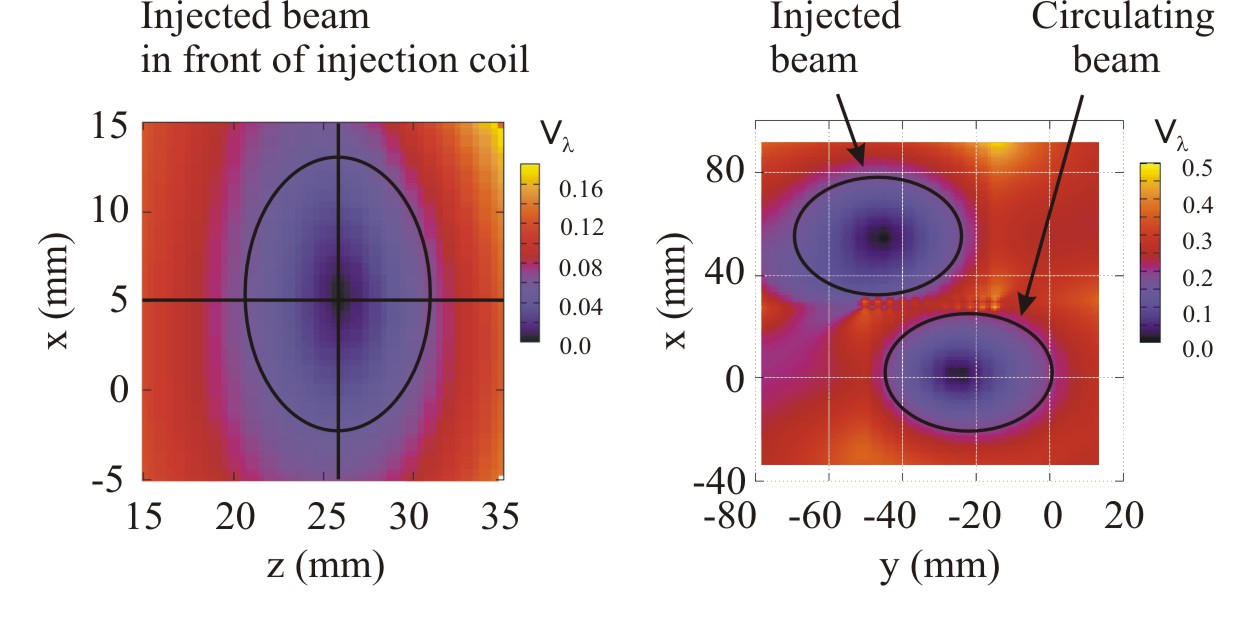}
\caption{\label{fig:13}Left: $x-z$ positions of the injected beam at the input colour coded with $v_\lambda$ at the output plane representing acceptance of the system. Right: Output $x-y$ distribution showing two beams.}
\end{figure}

Fig.~\ref{fig:13}  on the left shows the $v_{\lambda}$ mapping at the input plane $x-z$ i.e at the entrance of the injection coil.
It can be seen that $10~mm$ beam can be injected into the second toroid within acceptable pitch limit $v_{\lambda}<0.1$. Fig.~\ref{fig:13}  on the right shows the mapping at  the exit of Toroid-2.
Two beams can be seen. 
These two beams were simulated to match conditions at the output.

\section{\label{sec:level9}Conclusions and Outlook \protect\\
}
 
In this paper we have described the experimental activities under the project of  ``Figure-8 Storage Ring (F8SR)" for accumulation of intense low energy ion beams.
The experimental setup is very versatile for different beam transport experiments.
The ion beam transport in toroidal magnetic fields has been successfully demonstrated with \textit{in situ} detection method.
Further investigations would target to gain more knowledge about beam transport and to overcome the challenges poses by extreme conditions on beam diagnostics.

The main point will be experimental components operating in high magnetic field and high vacuum conditions.
The high magnetic field imposes limitations on electronics used for diagnostics e.g. digital camera, signal cables, inductors.
Moreover a magnetic shielding must be designed for any component that operates with rotating metal e.g. Rotary pumps.

The future experiment would stress on the optimal beam matching conditions.
The beam losses produce secondary electrons these can be trapped in the magnetic field.
The secondary electrons are observed to produce unwanted background structures influencing the beam signal.

In further steps, the magnetic field geometry will be tested by forming S-shape.
The drifts are expected to compensate due to the reversed direction of radius vector $\mathbf{R}$.
The experiments will be concluded by installation of the injection beam line and demonstration of ion beam injection in magnetic fields.


\end{document}